\documentclass[aip,preprint]{revtex4-1}
\usepackage{graphicx}
\usepackage[section]{placeins}
 \usepackage{float}
\draft 

\begin{document}


\title{Path integral molecular dynamics for thermodynamics and Green's function of ultracold spinor bosons}

\author{Yongle Yu}
\affiliation{Innovation Academy for Precision Measurement Science and Technology, Chinese Academy of Sciences, Wuhan 430071, P. R. China}

\author{Shujuan Liu}
\affiliation{College of Science, Zhejiang University of Technology, Hangzhou 31023, P. R. China}

\author{Hongwei Xiong}
\affiliation{College of Science, Zhejiang University of Technology, Hangzhou 31023, P. R. China}

\author{Yunuo Xiong}
\email{2111909023@zjut.edu.cn}
\affiliation{College of Science, Zhejiang University of Technology, Hangzhou 31023, P. R. China}

\begin{abstract}
Most recently, the path integral molecular dynamics has been successfully used to consider the thermodynamics of single-component identical bosons and fermions. In this work, the path integral molecular dynamics is developed to simulate the thermodynamics, Green's function and momentum distribution of two-component bosons in three dimensions. As an example of our general method, we consider the thermodynamics of up to sixteen bosons in a three-dimensional harmonic trap. For noninteracting spinor bosons, our simulation shows a bump in the heat capacity. As the repulsive interaction strength increases, however,  we find the gradual disappearance of the bump in the heat capacity. We believe this simulation result can be tested by ultracold spinor bosons with optical lattices and magnetic-field Feshbach resonance to tune the inter-particle interaction. We also calculate Green's function and momentum distribution of spinor bosons. Our work facilitates the exact numerical simulation of spinor bosons, whose property is one of the major problems in ultracold Bose gases.
\end{abstract}


\maketitle

\section{Introduction}

The exact numerical simulation of identical particles at finite temperature or zero temperature is one of the major problems in quantum physics. The so-called exact numerical simulation means that we should not assume in advance the trial wave function \cite{Rossi} or assume the usual contact interaction \cite{Parisi,Parisi1} for the convenience of calculations. The exact numerical simulation of the thermodynamics of identical particles requests that we carry out the theoretical studies directly from the partition function of the quantum system. Fortunately, without predetermining any properties of the quantum system, the partition function in the path integral formalism \cite{feynman,kleinert,Tuckerman} can be mapped to an equivalent classical ensemble of interacting ring polymers \cite{chandler,Parrinello,Miura,Cao,Cao2,Jang2,Ram,Kinugawa,Roy,Roy1,Roy2,Roy3,Kinugawa1,Kinugawa2,Roy4,Roy5,Roy6,Poly,Craig,Braa,Haber,Thomas}. However, during this mapping, one of the key obstacles we face is how to give exact consideration of quantum exchange effects for identical particles. In a pioneering work \cite{Hirshberg} on single-component identical bosons, a recursion formula is found to express the partition function in a highly efficient way. In Ref. \cite{Xiong}, path integral molecular dynamics (PIMD) and the recursion formula are generalized to calculate Green's function and momentum distribution of single-component bosons. Other recent advances include the application of PIMD to supersolid phase in high-pressure deuterium \cite{Deuterium} and bosons in singular magnetic field \cite{Xiong3}, and the generalization of PIMD to spin-polarized fermions \cite{HirshbergFermi,Xiong2,Xiong4}. 
In particular, PIMD was used most recently to consider the fermion sign problem for large fermion system \cite{Xiong4}.

Of course, we need to generalize PIMD to two-component (or spinor) bosons, because this is very common in quantum gases, in particular with intensive experimental and theoretical studies on ultracold spinor bosons \cite{Yuki,Mie,Sad,Chen,Wang}. The purpose of the present work is to develop PIMD for two-component bosons with the same intracomponent and intercomponent interaction, and it is straightforward to generalize our method to different intracomponent and intercomponent interactions. Different from previous PIMD simulation \cite{Hirshberg,Xiong,HirshbergFermi,Xiong2,Xiong3,Xiong4} for identical particles in two dimensions, here we consider spinor bosons in three dimensions and study the number of beads (in PIMD simulation) needed in numerical simulation to achieve numerical convergence. As an example, we consider the thermodynamics of two-component bosons in a three-dimensional harmonic trap. The energy and heat capacity are calculated to show the role of temperature and repulsive interaction in the thermodynamics of ultracold spinor bosons. We also generalize previous method \cite{Xiong,Xiong2} to calculate Green's function and momentum distribution of spinor bosons to pave the way of the general application of PIMD for ultracold spinor Bose gases.

\section{The partition function for two-component spinor bosons}

We consider in this work spinor bosons with two internal states denoted by $\left|\uparrow\right\rangle $ and $\left|\downarrow\right\rangle $.
It is straightforward to generalize to more general M-component spinor bosons. We consider
the following Hamiltonian operator which will not change the internal state of a boson:
\begin{equation}
\hat{H}=\sum_{j=1}^{N_\uparrow}\frac{\hat{\mathbf{p}}_{j}^{2}}{2m}+\sum_{j'=1}^{N_\downarrow}\frac{\hat{\mathbf{p}}_{j'}^{2}}{2m}+V\left(\hat{\mathbf{r}}_{1},\cdots,\hat{\mathbf{r}}_{N}\right).
\end{equation}
Here $N_\uparrow$ and $N_\downarrow$ denote the particle numbers in the internal states $|\uparrow>$ and $|\downarrow>$, respectively. $N=N_\uparrow+N_\downarrow$ is the total particle number of the quantum system. 
Here $V$ includes both the external potential
and inter-particle interaction.

At finite temperature $\beta=1/k_{B}T$, the partition function is
\begin{equation}
Z_B=Tr\left(e^{-\beta\hat{H}}\right).
\end{equation}
The purpose of the present work is to develop the PIMD to calculate the thermodynamics of spinor bosons based on this partition function.

For two-component spinor bosons, we define the following basis which is invariant about the exchange of any two bosons:
\begin{equation}
\left|N_B\right\rangle\sim\sum_{p\in S_N}\left|p\left\{\textbf{r}_{1}^{\uparrow},\cdots,\textbf{r}_{N_\uparrow}^{\uparrow};\textbf{r}_{N_\uparrow+1}^{\downarrow},\cdots,\textbf{r}_{N}^{\downarrow}\right\}\right\rangle .
\end{equation}
Here $p$ represents $N!$ permutation operations on
the index $\left\{ 1,2,\cdots,N_\uparrow,N_\uparrow+1,\cdots,N\right\} $. We do not change the order of the index for the internal
state in the permutation summation.

In this case, the partition function $Z_B$ for bosons is
\begin{equation}
Z_{B}\sim\int\prod_{j=1}^{N}d\textbf{r}_{j}\left\langle N_B\right|e^{-\Delta\beta\hat{H}}\cdots e^{-\Delta\beta\hat{H}}\left|N_B\right\rangle.
\end{equation}
Here $\Delta\beta=\beta/P$ and $P$ is the number of beads in PIMD.
One of the difficulties to calculate the partition function is the
$N!\times N!$ summation in the above expression.

Assume
\begin{equation}
\left|N\right\rangle =\left|\textbf{r}_{1}^{\uparrow},\cdots,\textbf{r}_{N_\uparrow}^{\uparrow};\textbf{r}_{N_\uparrow+1}^{\downarrow},\cdots,\textbf{r}_{N}^{\downarrow}\right\rangle
\end{equation}
without permutation summation, we have
\begin{equation}
Z_{B}\sim\int\prod_{j=1}^{N}d\textbf{r}_{j}\left\langle N\right|e^{-\Delta\beta\hat{H}}\cdots e^{-\Delta\beta\hat{H}}\left|N_B\right\rangle.
\label{Z1}
\end{equation}

Now we consider the following identity operator
\begin{equation}
\hat{I}_{B}\sim\int\prod_{j=1}^{N}d\textbf{r}_{j}\left|N_B\right\rangle\left\langle N_B\right|.
\end{equation}
By inserting a series of this identity operator in Eq. (\ref{Z1}), the partition function becomes:
\begin{equation}
Z_{B}\sim\int\prod_{j=1}^{N}d\textbf{r}_{j}\left\langle N\right|e^{-\Delta\beta\hat{H}}\hat{I}_{B}e^{-\Delta\beta\hat{H}}\cdots e^{-\Delta\beta\hat{H}}\hat{I}_{B}e^{-\Delta\beta\hat{H}}\hat{I}_{B}\left|N_B\right\rangle.
\end{equation}
Without considering the exchange symmetry, we also have another form
of the identity operator
\begin{equation}
\hat{I}=\int\prod_{j=1}^{N}d\textbf{r}_{j}\left|N\right\rangle \left\langle N\right|.
\end{equation}
In this case, it is easy
to prove that
\begin{equation}
e^{-\beta\hat{H}}\hat{I}_{B}=e^{-\frac{\beta}{P}\hat{H}}\hat{I}e^{-\frac{\beta}{P}\hat{H}}\hat{I}\cdots\hat{I}e^{-\frac{\beta}{P}\hat{H}}\hat{I}_{B}.
\end{equation}
We get
\begin{equation}
Z_{B}\sim\int\prod_{j=1}^{N}d\textbf{r}_{j}\left\langle N\right|e^{-\Delta\beta\hat{H}}\hat{I}e^{-\Delta\beta\hat{H}}\hat I\cdots \hat{I}e^{-\Delta\beta\hat{H}}\hat{I}e^{-\Delta\beta\hat{H}}\left|N_B\right\rangle.
\end{equation}

\begin{figure}[htbp]
\begin{center}
 \includegraphics[width=0.75\textwidth]{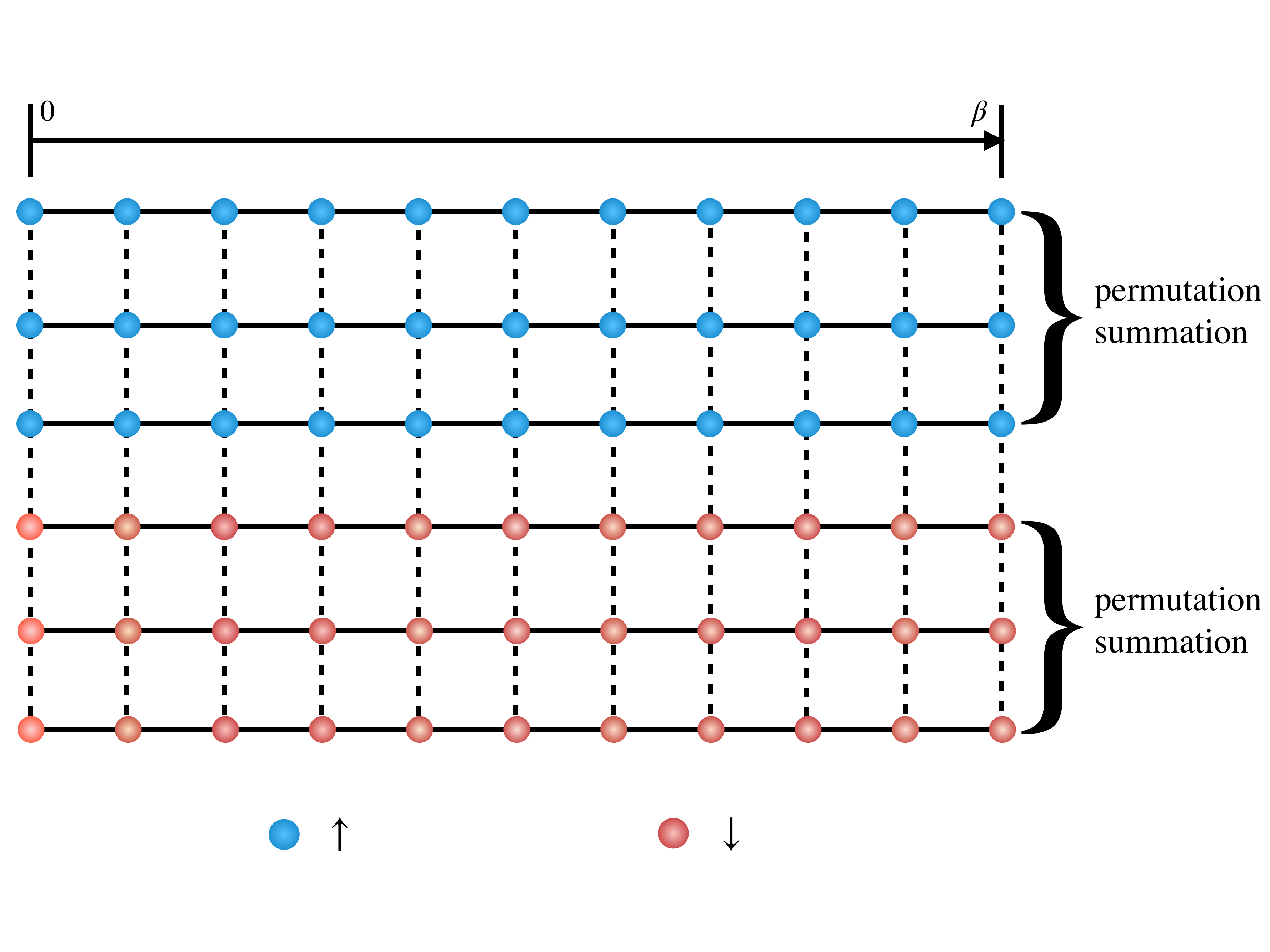} 
\caption{\label{Psummation} Illustration of the ring polymers for the partition function of spinor bosons with two internal states $|\uparrow>$ and $|\downarrow>$. The solid line shows the connection of the beads for the same particle, while the dashed line shows the inter-particle interaction between particles at the same imaginary time.}
\end{center}
\end{figure}

We see that for Hamiltonian operator which does not change the internal state of the bosons, in calculating
the partition function, we should consider all the permutations due to the
exchange symmetry for bosons in the same internal state. In Fig.
\ref{Psummation}, we illustrate an example of 3 bosons in internal state $\left|\uparrow\right\rangle $
and 3 bosons in internal state $\left|\downarrow\right\rangle $.
The dashed line in this figure illustrates the interaction between
particles, while the solid line shows the connection of the beads for the same particle.  
It is worthy to point out that in the integral $\int\prod_{j=1}^{N}d\textbf{r}_{j}$,
the coordinates are the same for $\left| N\right>$ and $\left|N_B\right\rangle$.
Hence, the permutation summation in $\left|N_B\right\rangle$ and closed ring polymer should be
considered simultaneously.

There are $N_\uparrow !\times N_\downarrow !$ permutation summations for the partition function shown in Fig. \ref{Psummation}, which increases exponentially with $N$. Fortunately, in a
recent pioneering work \cite{Hirshberg}, Hirshberg, Rizzi and Parrinello found
that the partition function corresponding to a system of identical
bosons without internal freedom can be calculated by a recursion formula which can be evaluated in $O(PN^{3})$ time, providing an
exponential speedup over simply summing all configurations for different
permutations. 

Generalizing the recursion formula for single-component bosons, we get the following expression of the partition function for two-component bosons
\begin{equation}
Z_{B}=\left(\frac{mP}{2\pi\hbar^{2}\beta}\right)^{PdN/2}\int d\textbf{R}_{1}...d\textbf{R}_{N}e^{-\beta U_{B}^{(N)}}.
\end{equation}
The system under consideration has $d$ spatial dimensions. Here $\mathbf{R}_{i}$
represents the collection of ring polymer coordinates $(\mathbf{r}_{i}^{1},...,\mathbf{r}_{i}^{P})$
corresponding to the $i$th particle. In addition,
\begin{equation}
U_{B}^{(N)}=-\frac{1}{\beta}\ln W_{B}^{(N_\uparrow)}-\frac{1}{\beta}\ln W_{B}^{(N_\downarrow)}+\frac{1}{P}\sum_{j=1}^{P}V\left(\textbf{r}_{1}^{j},...,\textbf{r}_{N}^{j}\right).
\end{equation}
\begin{equation}
W_{B}^{(N_\uparrow)}=\frac{1}{N_\uparrow}\sum_{k=1}^{N_\uparrow}e^{-\beta E_{N_\uparrow}^{(k)}}W_{B}^{(N_\uparrow-k)}.
\end{equation}

\begin{equation}
E_{N_\uparrow}^{(k)}=\frac{1}{2}m\omega_{P}^{2}\sum_{l=N_\uparrow-k+1}^{N_\uparrow}\sum_{j=1}^{P}\left(\textbf{r}_{l}^{j+1}-\textbf{r}_{l}^{j}\right)^{2}.
\end{equation}
Here $\textbf{r}_{l}^{P+1}=\textbf{r}_{l+1}^{1}$, except for $l=N_\uparrow$
for which $\textbf{r}_{N_\uparrow}^{P+1}=\textbf{r}_{N_\uparrow-k+1}^{1}$. In addition, $\omega_{P}=\sqrt{P}/\beta\hbar$.

\begin{equation}
W_{B}^{(N_\downarrow)}=\frac{1}{N_\downarrow}\sum_{k=1}^{N_\downarrow}e^{-\beta E_{N_\downarrow}^{(k)}}W_{B}^{(N_\downarrow-k)}.
\end{equation}

\begin{equation}
E_{N_\downarrow}^{(k)}=\frac{1}{2}m\omega_{P}^{2}\sum_{l=N_\downarrow-k+1}^{N_\downarrow}\sum_{j=1}^{P}\left(\textbf{r}_{N_\uparrow+l}^{j+1}-\textbf{r}_{N_\uparrow+l}^{j}\right)^{2}.
\end{equation}
Here $\textbf{r}_{N_\uparrow+l}^{P+1}=\textbf{r}_{N_\uparrow+l+1}^{1}$, except for
$l=N_\downarrow$ for which $\textbf{r}_{N}^{P+1}=\textbf{r}_{N-k+1}^{1}$.

The expression of $W_{B}^{(N_\uparrow)}$ ($W_{B}^{(N_\downarrow)}$) includes the exchange
effect by describing all the possible ring polymer configurations
for $N$ particles in the internal state $\left|\uparrow\right\rangle $
($\left|\downarrow\right\rangle $). The recursion formula for $W_{B}^{(N_\uparrow)}$ and $W_{B}^{(N_\downarrow)}$ is the same as that given in Ref. \cite{Hirshberg} for
bosons without considering internal state.  
We see that the
evaluation of the partition function still takes $O(PN^{3})$ time, rather than $O(PN_\uparrow^{3} N_\downarrow^3)$. 
For two-component spinor bosons there are in total $N_\uparrow ! N_\downarrow !$ terms in the partition function summation for different configurations. However, because spin degree of freedom doesn’t change, the spin up contribution and spin down contribution to the partition function are unrelated, and the partition function can be divided into two parts and evaluated separately, so the evaluation of the partition function presented in this work takes $O(PN_\uparrow^3+PN_\downarrow^3)$ times, which is still $O(PN^3)$. However, if the spin degree of freedom has its own dynamics due to spin-changing interaction, then there is no such shortcut and the evaluation of the full partition function now takes $O(P N_\uparrow^3 N_\downarrow^3)$ times.

After we get the final expression of $Z_{B}$, we may use massive
Nos\'e-Hoover chain \cite{Nose1,Nose2,Hoover,Martyna,Jang} to establish constant temperature
for the system, where each degree of freedom of the system has been
coupled to a separate Nos\'e-Hoover thermostat. From
the distribution generated by this method, we can calculate the density
distribution and energy of the system.  For example, the average density is simply given by
\begin{equation}
\rho_\uparrow(\textbf{x})=\left<\frac{1}{P}\sum_{j=1}^P\sum_{k=1}^{N_\uparrow}\delta(\mathbf r_k^j-\textbf{x})\right>,
\rho_\downarrow(\textbf{x})=\left<\frac{1}{P}\sum_{j=1}^P\sum_{k=N_\uparrow+1}^{N}\delta(\mathbf r_k^j-\textbf{x})\right>.
\end{equation}
The energy estimator can be obtained from the following expression:
\begin{equation}
E(\beta)=-\frac{1}{Z_B}\frac{\partial Z_B}{\partial\beta}.
\end{equation}
The energy estimator is
\begin{equation}
E(\beta)=\frac{PdN}{2\beta}+\frac{\left<U\right>}{P}+\left<V_B^{(N_\uparrow)}+\beta\frac{\partial V_B^{(N_\uparrow)}}{\partial\beta}\right>+\left<V_B^{(N_\downarrow)}+\beta\frac{\partial V_B^{(N_\downarrow)}}{\partial\beta}\right>.
\end{equation}
where $V_B^{(N_\uparrow)}=-\frac{1}{\beta}\log W_{B}^{(N_\uparrow)}$ and $V_B^{(N_\downarrow)}=-\frac{1}{\beta}\log W_{B}^{(N_\downarrow)}$. $V_B^{(N_\uparrow)}+\beta\frac{\partial V_B^{(N_\uparrow)}}{\partial\beta}$ may be evaluated as
\begin{equation}
V_B^{(N_\uparrow)}+\beta\frac{\partial V_B^{(N_\uparrow)}}{\partial\beta}=\frac{\sum_{k=1}^{N_\uparrow}\left[V_B^{(N_\uparrow-k)}+\beta\frac{\partial V_B^{(N_\uparrow-k)}}{\partial\beta}-E_{N_\uparrow}^{(k)}\right]e^{-\beta(E_{N_\uparrow}^{(k)}+V_B^{(N_\uparrow-k)})}}{\sum_{k=1}^{N_\uparrow} e^{-\beta(E_{N_\uparrow}^{(k)}+V_B^{(N_\uparrow-k)})}},
\end{equation}
with $V_B^{(0)}+\beta\frac{\partial V_B^{(0)}}{\partial\beta}=0$. Similar formulas hold for $V_B^{(N_\downarrow)}+\beta\frac{\partial V_B^{(N_\downarrow)}}{\partial\beta}$. The interaction potential $U$ is given by
\begin{equation}
U=\sum_{j=1}^{P}V\left(\textbf{r}_{1}^{j},...,\textbf{r}_{N}^{j}\right).
\end{equation}
One may refer to our open-source code with C++ to see the details of the implementation for the simulation of the partition function, energy estimator and density estimator, etc..

\section{Results for two-component spinor bosons}

In our PIMD simulation, the number of beads used decreases as temperature increases to ensure numerical stability and assure convergence. Compared with two dimensions with $P=12/T$ beads in the usual case \cite{Xiong,Xiong2,Xiong3}, it is worthy to point out that many more beads are needed to assure accuracy and convergence, because the size of phase space has been greatly enlarged compared with two dimensional case. Based on the relation $\Delta\beta=\beta/P$ and the fact that the evolution operator has been expanded to an order of $O(\Delta\beta^2)$, we would like to keep $\Delta\beta$ a constant with different $\beta$, which leads to the choice P=70/T. In general, of course, the convergence with respect to the number of beads has to be explicitly verified by increasing the number of beads and check if the results remain the same.  It’s true that in this way one finds that non-interacting case requires less beads than the interacting case, but we choose the same number of beads for all cases while ensuring that the results converge with respect to the number of beads. Moreover, we performed 12 independent simulations each with at least $10^6$ MD steps to establish thermal equilibrium and calculate expectation values.
In all of the following we checked convergence with respect to the number of beads and MD steps performed, by increasing P and MD steps and check if we obtain the same results. In our simulation, the statistical fluctuations are negligible. Hence, there is no error bar in all figures for numerical results.

\subsection{Noninteracting spinor Bose gas in 3D harmonic trap}

\begin{figure}[htb]
\begin{center}
 \includegraphics[width=0.75\textwidth]{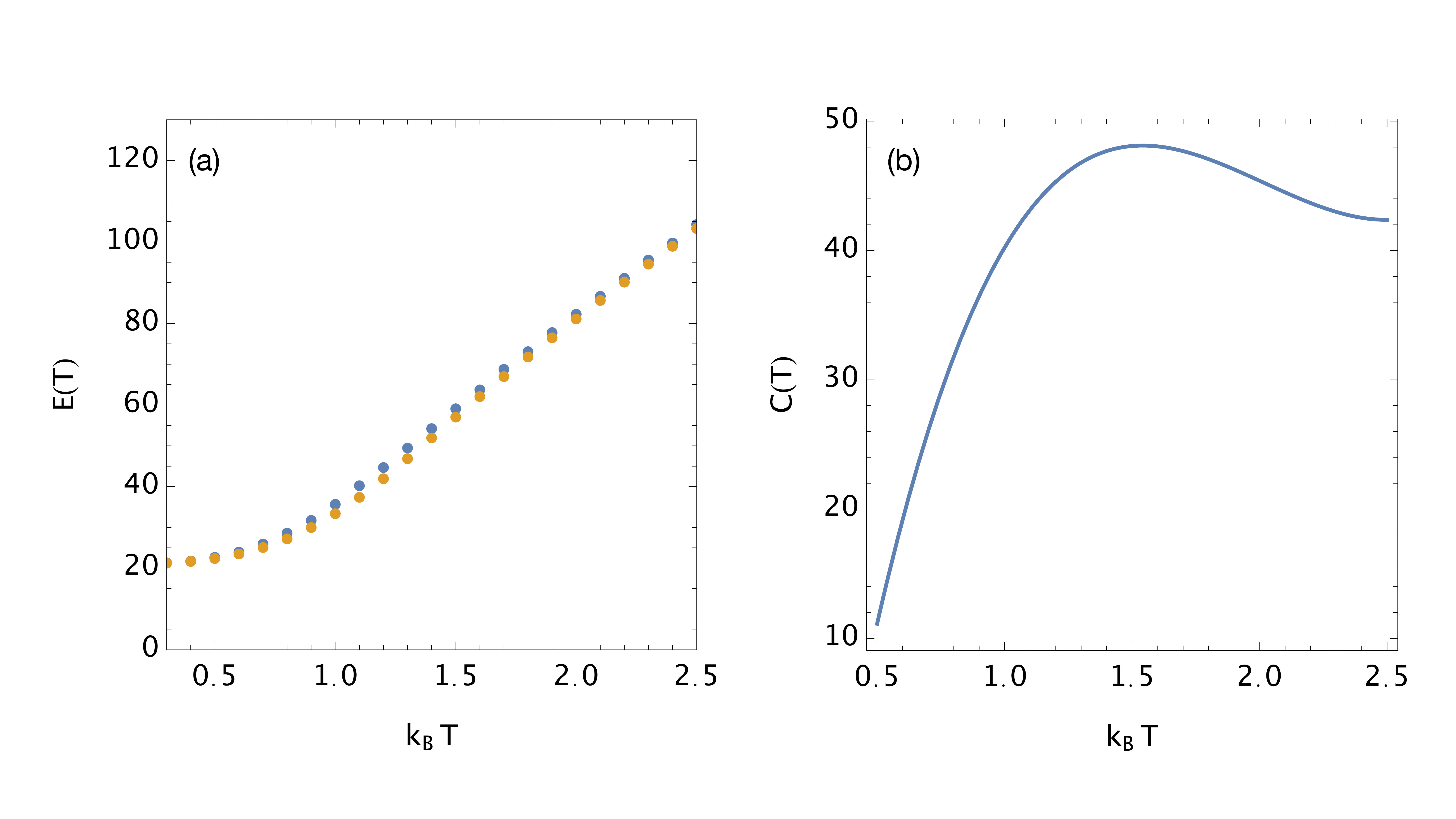} 
\caption{\label{averageE} For $N_{\uparrow}=N_{\downarrow}=7$ non-interacting bosons in three-dimensional harmonic trap, (a) shows the average energy by PIMD for spinor bosons (orange circle) and the average energy calculated by grand canonical ensemble (blue circle). The good agreement proves the validity and accuracy of our PIMD simulation. (b) shows the heat capacity calculated from the average energy of PIMD, which shows a clear bump as the temperature decreases. The statistical fluctuations are negligible.
 }
\end{center}
\end{figure}

To test our algorithm, we first consider noninteracting identical bosons for $N_{\uparrow}=N_{\downarrow}=7$
in a three-dimensional harmonic trap with angular
frequency $\omega$. We use the natural units of $\hbar=1,m=1$ and
consider $\omega=1$. The dimensionless Hamiltonian operator is
\begin{equation}
\hat{H}=\frac{1}{2}\sum_{j=1}^{N}\hat{\mathbf{p}}_{j}^{2}+\frac{1}{2}\sum_{j=1}^{N}\hat{\textbf{r}}_{j}^{2}.
\end{equation}
In Fig. \ref{averageE}(a), the orange circle gives the average energy in our numerical simulation for different temperatures, while the blue circle gives the average energy in grand canonical ensemble for ideal bosons. In this work, the heat capacity is calculated by the finite-difference derivative for the energy with respect to the temperature. In Fig. \ref{averageE}(b), we give the heat capacity $C(T)$ based on the average energy calculated by PIMD. Our simulation shows the presence of a bump in the heat capacity. In the thermodynamic limit, the critical temperature \cite{Bagnato,Pitaevskii} is determined by $k_BT_c=0.94\hbar(\omega_x\omega_y\omega_z)^{1/3}N_\uparrow^{1/3}$. We have $T_c\approx 1.8$ with our parameters. The peak in our simulation for the heat capacity is at $T_c\approx 1.53$, which is smaller than the result with the formula of the thermodynamic limit. This comparison agrees qualitatively with the result that the finite-size effect has the effect of decreasing the critical temperature \cite{Shujuan,Xiong-finite}.

\begin{figure}[H]
\begin{center}
 \includegraphics[width=0.75\textwidth]{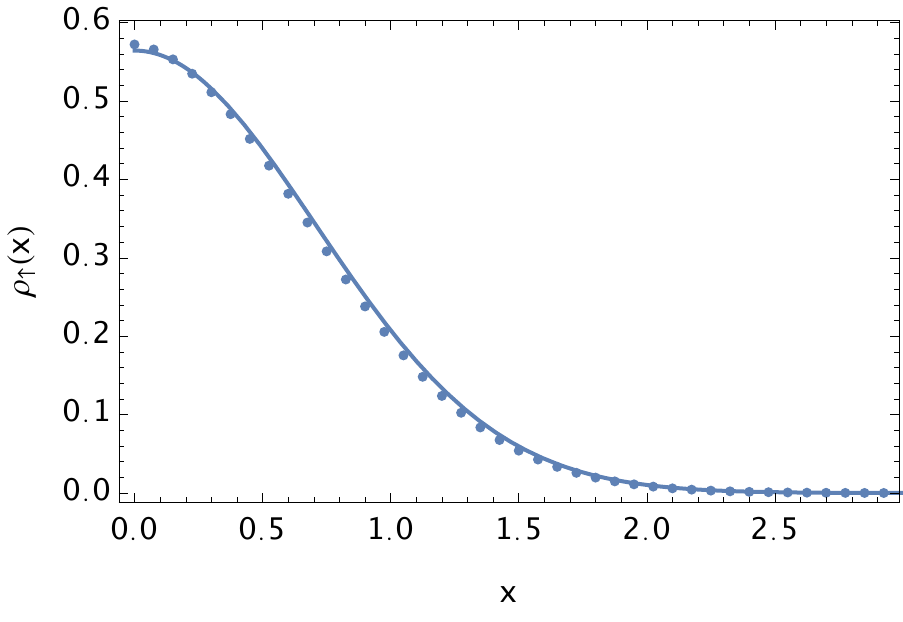} 
\caption{\label{density} For $N_{\uparrow}=N_{\downarrow}=7$ non-interacting bosons in a three-dimensional harmonic trap at the temperature $k_BT=0.3$, the circle shows the density distribution $\rho_\uparrow(x)$ by PIMD, while the solid line is the analytical result of the ground state density distribution. The statistical fluctuations are negligible.
 }
\end{center}
\end{figure}

In Fig. \ref{density}, for $k_BT=0.3$ (in unit of $\hbar\omega$) we give the
numerical result of the normalized density distribution $\rho_{\uparrow}(x)$ along $x$ direction by
solid circle, while the solid line gives the analytical result of
the ground state. For $k_{B}T<<\hbar\omega$, the density distribution
has the following normalized analytical approximation:
\begin{equation}
\rho_{\uparrow}(x)\simeq \left(\frac{m\omega}{\pi\hbar}\right)^{1/2}e^{-\frac{m\omega}{\hbar }x^{2}}.
\end{equation}
Good agreement is found with our simulation results.

\subsection{Spinor Bose gas in three-dimensional harmonic trap with repulsive Gaussian interaction}

The great merit of PIMD for identical bosons lies in that the inclusion of inter-particle interaction will not add too much additional difficulty and calculation error.

We consider the following Gaussian interaction \cite{Mujal} between
bosons
\begin{equation}
V(\mathbf{r}_{1},...,\mathbf{r}_{N})=\frac{1}{2}\sum_{j\neq j'=1}^N\frac{g}{\gamma^2\pi} e^{-|{\textbf r}_j-{\textbf r}_{j'}|^2/\gamma^2}.
\label{Vinteraction}
\end{equation}

\begin{figure}[H]
\begin{center}
 \includegraphics[width=0.75\textwidth]{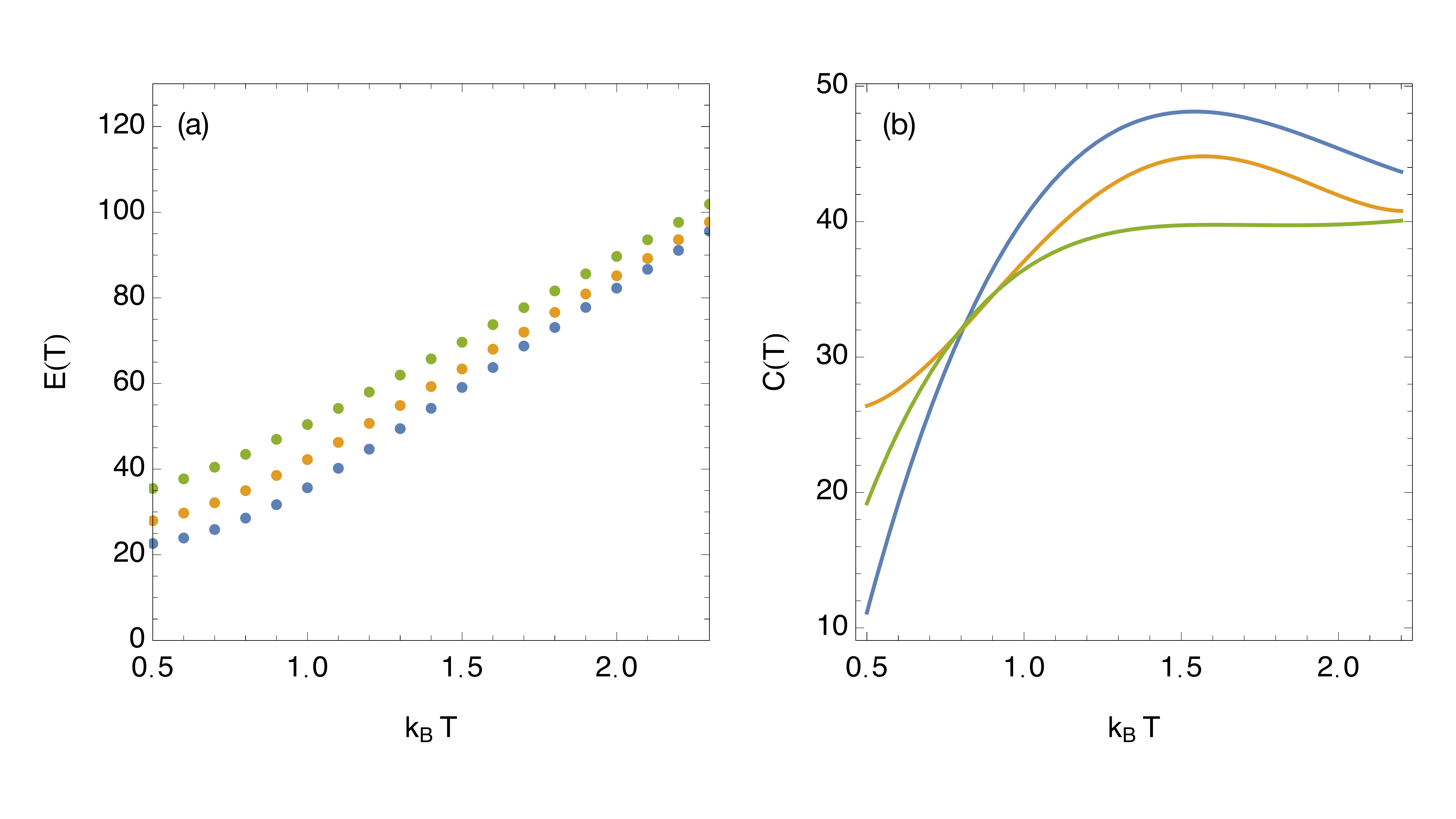} 
\caption{\label{intG} For $N_{\uparrow}=N_{\downarrow}=7$ bosons in a three-dimensional harmonic trap with $g=0$ (blue), $g=1$ (orange) and $g=3$ (green), (a) shows the average energy for different $g$, while (b) shows the heat capacity for different $g$. As the inter-particle coupling constant increases, we see a gradual disappearance of the bump in the heat capacity. The statistical fluctuations are negligible.
 }
\end{center}
\end{figure}

We consider here $\gamma=1$. With length unit $\sqrt{\hbar/m\omega}$ and energy unit $\hbar\omega$, we consider the average energy for $N_\uparrow=N_\downarrow=7$ with different coupling constant $g$. In Fig. \ref{intG}(a), we show the average energy for different $g$, while in Fig. \ref{intG}(b) the heat capacity for different $g$ is given. As expected, at the same temperature, the repulsive interaction has the effect of increasing the total energy. Significant role of the inter-particle repulsive interaction is shown by the heat capacity, which is still quite challenging to reveal in mean-field theory or the theory beyond the mean-field theory. Our accurate simulations show that the repulsive inter-particle interaction will gradually erase the bump in the heat capacity of ideal bosons. This unique result will provide the benchmark for other methods in future to calculate accurately the thermodynamics of interacting bosons. The disappearance of the bump in the heat capacity is due to the strong interaction between particles.

We consider here Gaussian interaction, because for single-component bosons with Gaussian interaction in two dimensions, an analytical expression is given by Mujal et al. \cite{Mujal} for $N\leq 4$ bosons. Our simulation may give the benchmark for other method to calculate accurately the energy of spinor bosons in three dimensions in future. The s-wave scattering length $a$ may estimated with $\int d^3 \textbf r \frac{g}{\gamma^2\pi} e^{-|\textbf r|^2/\gamma^2}=4\pi\hbar^2 a/m$ \cite{Pitaevskii}. In this case, we have the s-wave scattering length $a=\frac{g\gamma}{4\sqrt {\pi}}$. It is worth pointing out that, to compare with experiments of cold bosonic atoms with s-wave scattering length $a$, the Gaussian interaction should be chosen to satisfy $a/\bar l<<1$, with $\bar l$ the average interparticle distance.

\begin{figure}[H]
\begin{center}
 \includegraphics[width=0.75\textwidth]{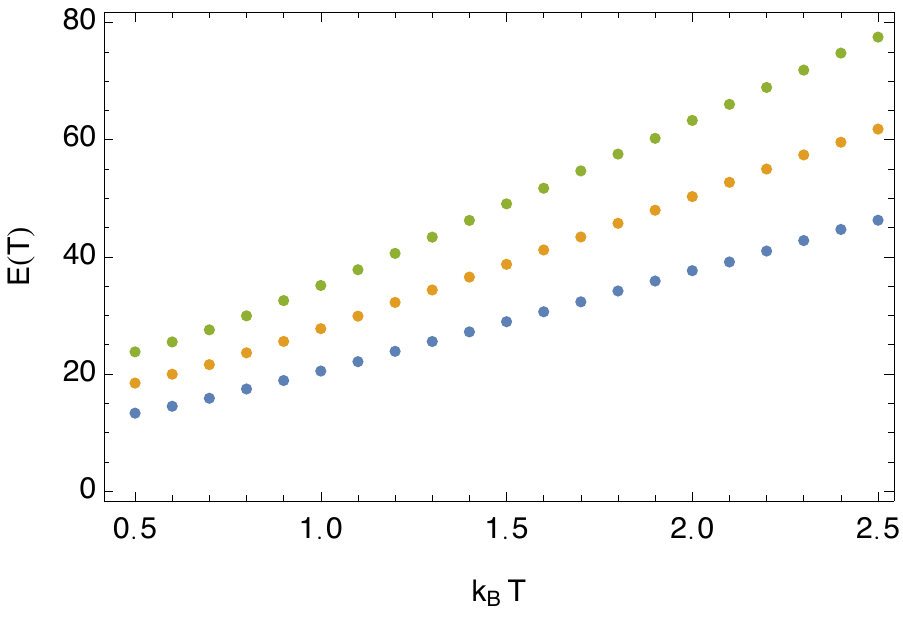} 
\caption{\label{DifferentP} The simulated energy for  $N_\uparrow=N_\downarrow=$ 3 (blue), 4 (orange), 5 (green) with the coupling constant $g=3$. The statistical fluctuations are negligible.}
\end{center}
\end{figure}

In Fig. \ref{DifferentP}, we give the simulation result of the average energy for $N_\uparrow=N_\downarrow=3,4,5$ with the coupling constant $g=3$.

\subsection{Spinor Bose gas in three-dimensional harmonic trap with dipole interaction}

Now we consider dipole interaction which is the real situation of spinor Bose gases \cite{Yuki}. We consider the following simplest dipole interaction 
\begin{equation}
V(\mathbf{r}_{1},...,\mathbf{r}_{N})=\frac{1}{2}\sum_{j\neq j'=1}^N\frac{\lambda}{|{\textbf r}_j-{\textbf r}_{j'}|^3} .
\label{dipole}
\end{equation}
Usually, the realistic dipole interaction comprises of an anisotropic term \cite{Stuhler}. The above dipole interaction is also used in path integral Monte Carlo simulations of fermions \cite{Dornheim}. We use this dipole interaction so that it has the chance to provide the benchmark for other relevant studies.

\begin{figure}[H]
\begin{center}
 \includegraphics[width=0.75\textwidth]{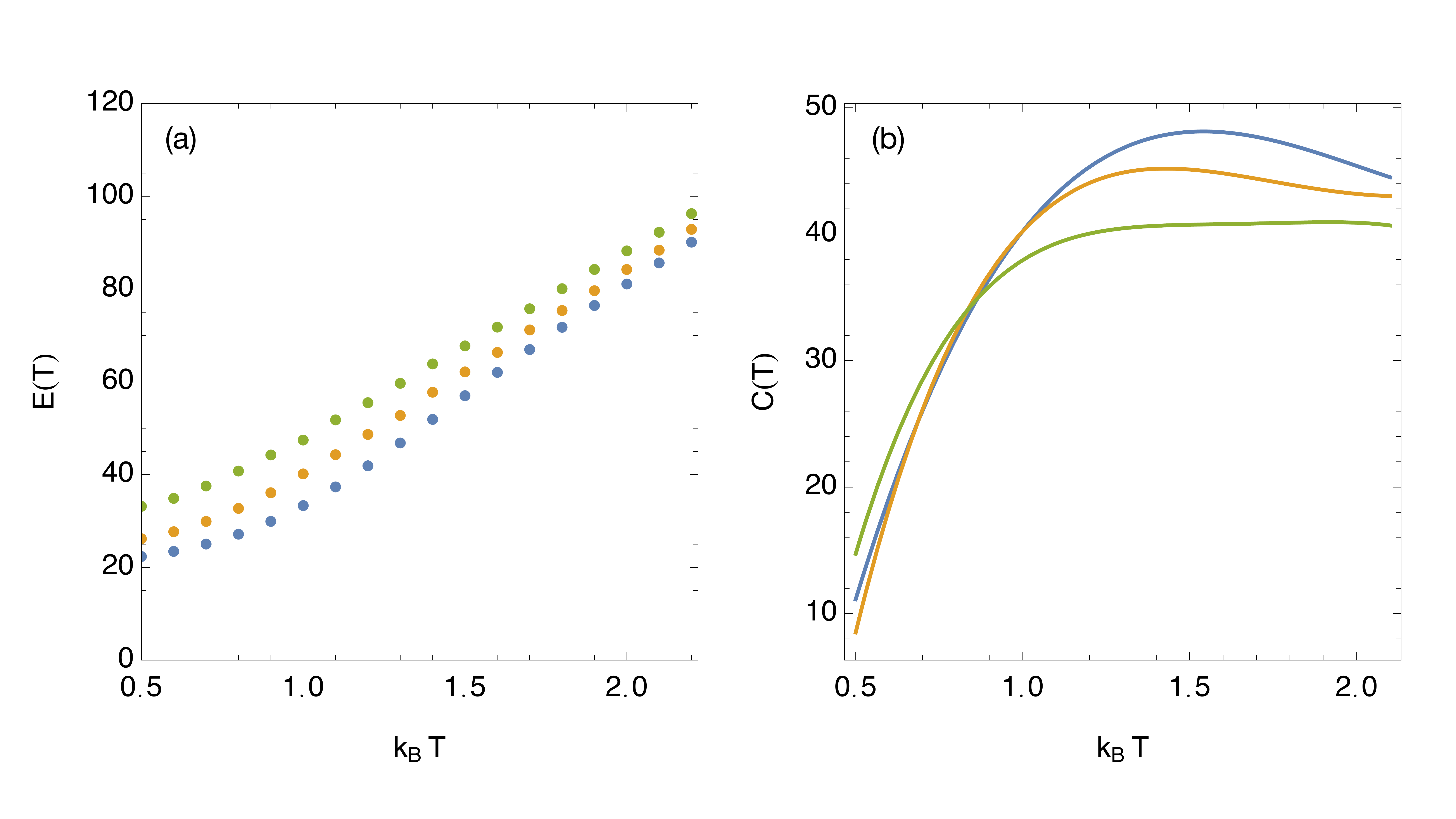} 
\caption{\label{DifferentL} For $N_{\uparrow}=N_{\downarrow}=7$ bosons in a three-dimensional harmonic trap with dipole interaction, we show the result of different coupling constant $\lambda=0$ (blue), $\lambda=0.02$ (orange) and $\lambda=0.1$ (green). (a) shows the average energy for different $\lambda$, while (b) shows the heat capacity for different $\lambda$. As the inter-particle coupling constant increases, we see a gradual disappearance of the bump in the heat capacity, which is similar to the case of Gaussian interaction. The statistical fluctuations are negligible.
 }
\end{center}
\end{figure}

By varying the interaction strength $\lambda$ we observed similar phenomenon as we did with Gaussian interaction, as shown in  Fig. \ref{DifferentL}.

\section{Green's function and momentum distribution}

Now we turn to consider the momentum distribution of spinor bosons. In our previous work, we calculated the momentum distribution for single-component bosons in PIMD by first considering Green's function of the system. Our method can be generalized directly to the case of spinor bosons.

For spinor bosons, the thermal Green's
function is defined as 
\begin{equation}
G_{\uparrow\uparrow}(\textbf{x},\tau_{1};\textbf{y},\tau_{2})=\left<\mathcal{T}\left\{ \hat{\psi}_{\uparrow}(\mathbf{y},\tau_{2})\hat{\psi}_{\uparrow}^{\dagger}(\mathbf{x},\tau_{1})\right\} \right>,
\end{equation}
where $\left<\cdots\right>$ denotes thermal average, $\mathcal{T}$
is the imaginary time-ordering operator. In addition, 
\begin{equation}
\hat{\psi}_{\uparrow}(\mathbf{x},\tau)=e^{\hat{H}\tau}\hat{\psi}_{\uparrow}(\mathbf{x})e^{-\hat{H}\tau},\hat{\psi}^{\dagger}_{\uparrow}(\mathbf{x},\tau)=e^{\hat{H}\tau}\hat{\psi}^{\dagger}_{\uparrow}(\mathbf{x})e^{-\hat{H}\tau}.
\end{equation}
Of course, we have $G_{\uparrow\downarrow}=0$ and $G_{\downarrow\uparrow}=0$.

To get momentum distribution, we should calculate
\begin{equation}
G_{\uparrow\uparrow}(\textbf{x},\tau_{2}+\Delta\beta;\textbf{y},\tau_{2})=\left<\hat{\psi}_{\uparrow}^{\dagger}(\mathbf{x},\tau_{2}+\Delta\beta)\hat{\psi}_{\uparrow}(\mathbf{y},\tau_{2})\right>.
\end{equation}

\begin{figure}[htbp]
\begin{center}
 \includegraphics[width=0.75\textwidth]{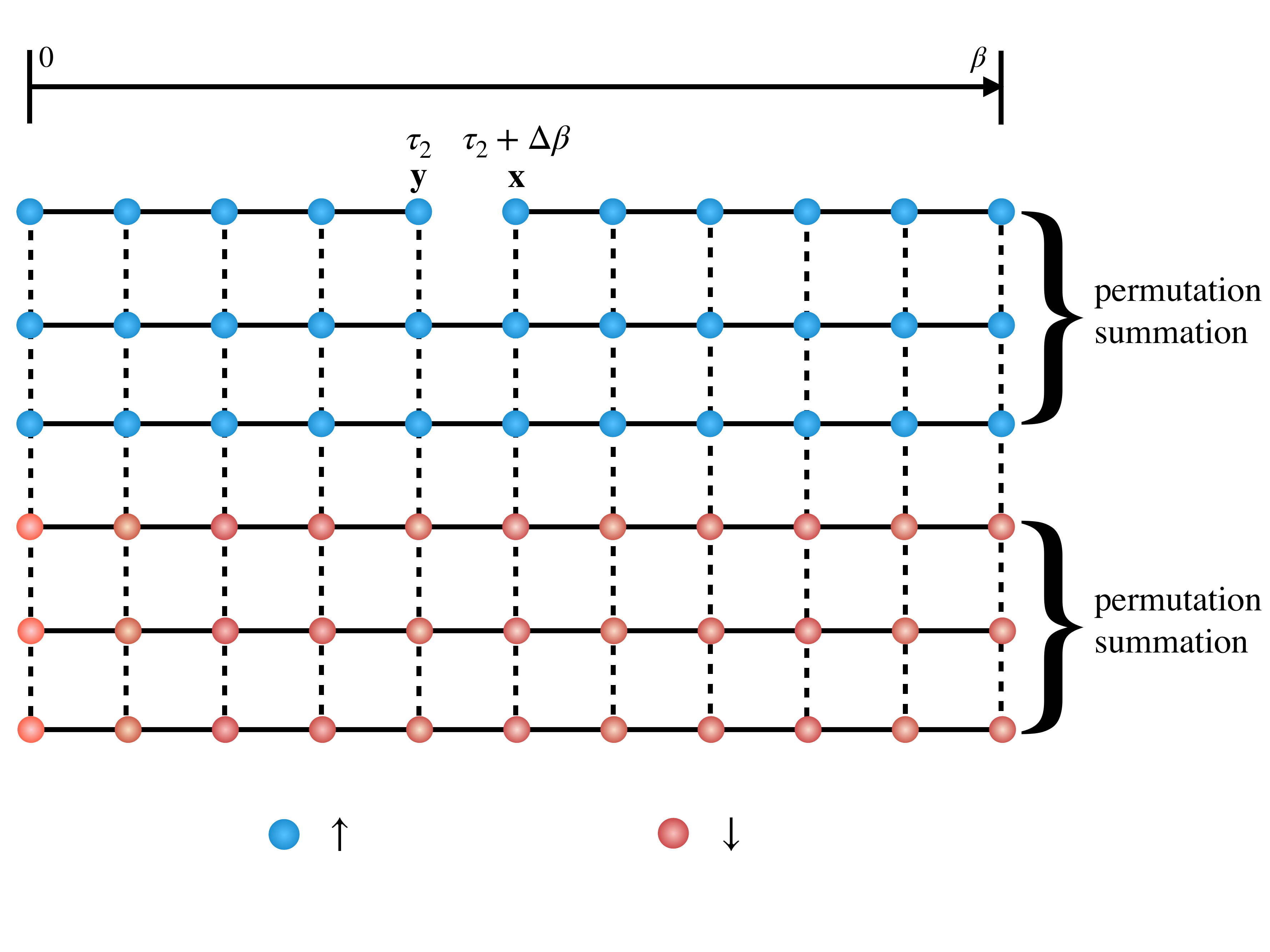} 
\caption{\label{GreenIllustration} Modified ring polymers for Green's function $G_{\uparrow\uparrow}$. There is a gap between two beads, from $\tau_2$ to $\tau_2+\Delta\beta$. In particular, the coordinates of the beads at the ends of the gap are $\mathbf y$ and $\mathbf x$. There is no harmonic spring between those two beads.}
\end{center}
\end{figure}

In Fig. \ref{GreenIllustration}, we show the idea to calculate this Green's function. One may find the details to implement the algorithm in our previous work. Green's function can be estimated as
\begin{equation}
G_{\uparrow\uparrow}(\textbf{x}',\tau_1;\textbf{y}',\tau_2)=\left<\delta(\textbf{x}-\textbf{x}')\delta(\textbf{y}-\textbf{y}')\right>,
\end{equation}
where $\textbf{x}$ and $\textbf{y}$ denote the positions of two beads at the end of the gap in Fig. \ref{GreenIllustration}.

As illustrated in Fig. \ref{GreenIllustration}, there would be a gap for the beads in the
top row, due to the field operator $\hat{\psi}_{\uparrow}^{\dagger}(\mathbf{x},\tau_{2}+\Delta\beta)\hat{\psi}_{\uparrow}(\mathbf{y},\tau_{2})$.
Following the method given in Ref. \cite{Xiong}, it is straightforward to write
the recursion formula for this Green's function. 

Based on thermal Green's function, the momentum distribution is
\begin{equation}
\rho_{\uparrow}(\textbf{p})=\frac{1}{(2\pi\hbar)^{d}}\int d\textbf{x}d\textbf{y}G_{\uparrow\uparrow}(\textbf{x},\tau_{2}+\Delta\beta;\textbf{y},\tau_{2})e^{\frac{i}{\hbar}\textbf{p}\cdot(\textbf{x}-\textbf{y})}.
\end{equation}
It is similar to consider $G_{\downarrow\downarrow}(\textbf{x},\tau_{2}+\Delta\beta;\textbf{y},\tau_{2})$
and $\rho_{\downarrow}(\textbf{p})$.

\begin{figure}[H]
\begin{center}
 \includegraphics[width=0.75\textwidth]{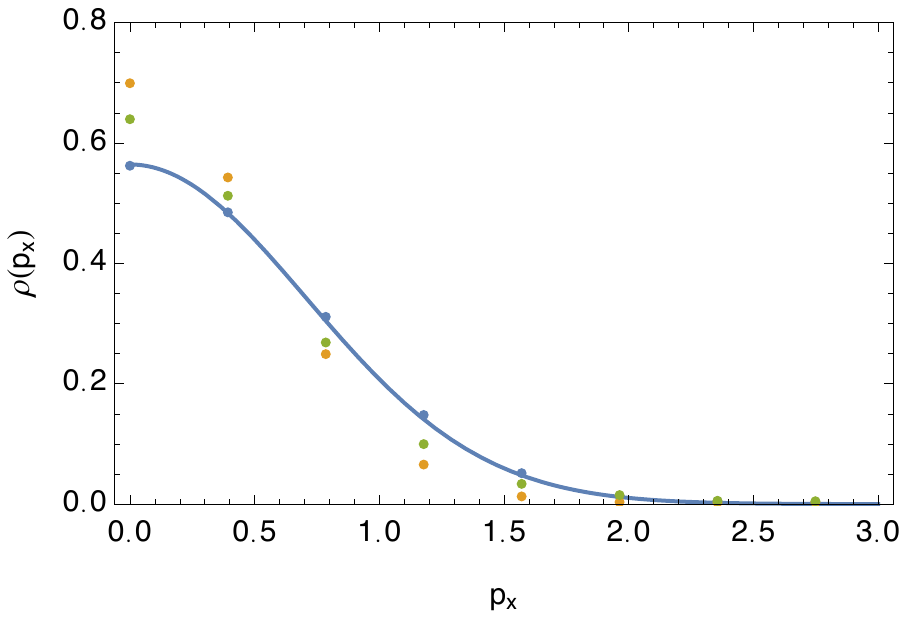} 
\caption{\label{momentum} Shown is the momentum distribution $\rho(p_x)$ of $N_\uparrow=N_\downarrow=8$ bosons in 3D harmonic trap with different dipole interaction at $k_BT=0.3$. The blue circle is the momentum distribution of $\lambda=0$, while the green circle and orange circle are for $\lambda=0.5$ and $\lambda=1$, respectively. The solid line is the momentum distribution of the ground state wave function, which agrees well with our simulation at low temperature. The statistical fluctuations are negligible.}
\end{center}
\end{figure}

As an application of our method, we consider $N_\uparrow=N_\downarrow=8$ bosons in 3D harmonic trap with different dipole interaction at $k_BT=0.3$. In Fig. \ref{momentum}, we show the momentum distribution along $x$ direction for different $\lambda$. We see that as $\lambda$ increases, the momentum distribution becomes narrower, as expected from the physical picture that repulsive interactions have the effect of increasing the width of the quantum system. 
At the same temperature, this effect of inter-particle interaction for momentum distribution may be also interpreted by considering Green's function in real space. The ground state wave function is a Gaussian function for non-interacting case, while in the Thoms-Fermi approximation the quadratic behavior is dominating as interaction increases so that it becomes flatter \cite{Pitaevskii}.
In experiments, one may use optical lattices \cite{lattice} so that in a lattice site there exists a few or a dozen spinor bosons. By using magnetic-field Feshbach resonance \cite{Feshbach} to turn the inter-particle interaction, we have the chance to show experimentally the behavior of the heat capacity simulated in this work. 

\section{Conclusion and discussion}

As a summary, in this work we showed how to generalize the original PIMD methodology to consider spinor bosons and extract thermal properties from our simulations, such as density, heat capacity and momentum distribution. By using our exact numerical method, we are able to study the behavior of heat capacity and we observed that as inter-particle interaction strength increases, the bump in heat capacity curve is gradually erased. Issues worth considering in the future are the compressibility in a harmonic trap, which has been calculated by path integral Monte Carlo for a homogeneous
dilute Bose gas and a discontinuous behavior is found at the critical point \cite{Spada}. We also suggest possible experiment scheme to observe this behavior experimentally. Of course, the method we developed here can be generalized straightforwardly to consider bosons with multiple components and other systems, providing opportunities for future research.

\begin{acknowledgments}
This work
is partly supported by the National Natural Science Foundation of China under grant numbers 11175246, and 11334001. 
\end{acknowledgments}

\textbf{DATA AVAILABILITY}

The data that support the findings of this study are available from the corresponding author upon reasonable request.
The code of this study is openly available in GitHub (https://github.com/xiongyunuo/PIMD-Pro-SB-MT).

\end{document}